\documentclass{moriond}

\usepackage{amsfonts}
\usepackage{tikz}
\usepackage{amsmath}
\usetikzlibrary{arrows.meta, positioning, calc}

\def\be{\begin{equation}}
\def\ee{\end{equation}}
\def\bea{\begin{eqnarray}}
\def\eea{\end{eqnarray}}

\newcommand{\Photo}{}

\begin{document}
\vspace*{4cm}

\title{Beyond AI as Assistants: Toward Autonomous Discovery in Cosmology}





\author{Licong Xu$^{1,2}$ and Thomas Borrett$^{2,3}$}

\address{$^1$ Institute of Astronomy, University of Cambridge, Madingley Road, Cambridge CB3 0HA, UK\\
$^2$ Kavli Institute for Cosmology, University of Cambridge, Madingley Road, Cambridge CB3 0HA, UK\\
$^3$ Cavendish Astrophysics, University of Cambridge, Madingley Road, Cambridge CB3 0HA, UK}


\maketitle\abstracts{
Recent advances in artificial intelligence (AI) agents are pushing AI beyond tools toward autonomous scientific discovery. We discuss two complementary agentic systems for cosmology: \texttt{CMBEvolve}, which targets tasks with explicit quantitative objectives through LLM-guided code evolution and tree search, and \texttt{CosmoEvolve}, which targets open-ended scientific workflows through a virtual multi-agent research laboratory. As preliminary demonstrations, we apply \texttt{CMBEvolve} to out-of-distribution detection in weak-lensing maps, where it iteratively improves the benchmark score through code evolution, and \texttt{CosmoEvolve} to exploratory ACT DR6 data analysis, where it produces map-level diagnostics and identifies pair- and scale-dependent patterns that require expert interpretation within the full ACT analysis framework. These examples show how cosmology can provide both controlled benchmark tasks and realistic open-ended research problems for the development of AI scientist systems.}

\vspace{-1.0cm}   
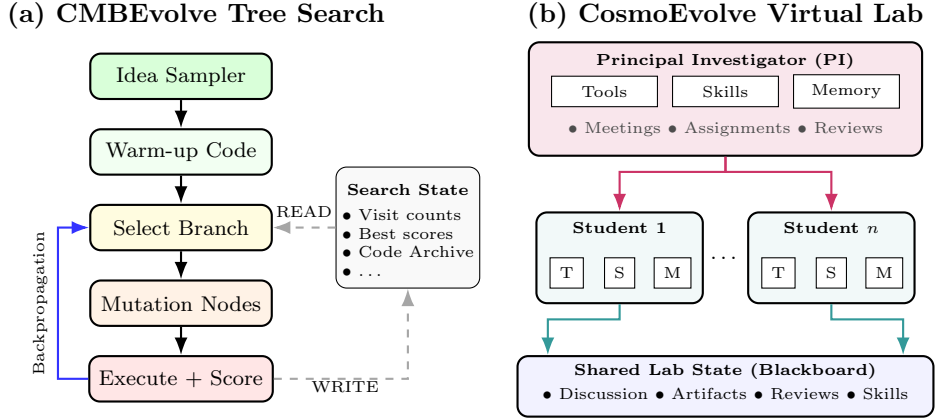
\begin{figure}[ht]
\centering
\begin{tikzpicture}[
    >=Latex,
    font=\scriptsize, 
    every node/.style={align=center},
    box/.style={draw, rounded corners, thick, minimum width=2.4cm, minimum height=0.6cm},
    ideas/.style={box, fill=green!15},
    warmup/.style={box, fill=green!5},
    select/.style={box, fill=yellow!15},
    mutate/.style={box, fill=orange!10},
    eval/.style={box, fill=red!10},
    flow/.style={->, thick},
    backprop/.style={->, thick, color=blue!80},
    stateflow/.style={->, thick, dashed, color=gray!70}
]

\node[font=\small\bfseries] (titleA) at (0, 0.8) {(a) CMBEvolve Tree Search};

\node[ideas]  (ideas)  at (-0.2, 0)      {Idea Sampler};
\node[warmup] (warmup) at (-0.2, -1)   {Warm-up Code};
\node[select] (select) at (-0.2, -2)   {Select Branch};
\node[mutate] (mutate) at (-0.2, -3)   {Mutation Nodes};
\node[eval]   (eval)   at (-0.2, -4)   {Execute + Score};

\draw[flow] (ideas) -- (warmup);
\draw[flow] (warmup) -- (select);
\draw[flow] (select) -- (mutate);
\draw[flow] (mutate) -- (eval);

\draw[backprop] (eval.west) -- ++(-0.4,0) |- (select.west) 
    node[pos=0.25, rotate=90, anchor=south, color=black, font=\tiny] {Backpropagation};

\node[draw, rounded corners, fill=gray!5, minimum width=1.9cm, minimum height=1.6cm] (state) at (2.8, -2) {};
\node[font=\tiny\bfseries] at (2.8, -1.5) {Search State};
\node[font=\tiny, align=left] at (2.8, -2.2) {
    $\bullet$ Visit counts\\
    $\bullet$ Best scores\\
    $\bullet$ Code Archive\\
    $\bullet$ \dots
};

\draw[stateflow] (state.west) -- (select.east) 
    node[midway, above, font=\tiny, color=black] {READ};
\draw[stateflow] (eval.east) -| (state.south) 
    node[pos=0.10, right, font=\tiny, color=black, yshift=-3pt] {WRITE};

\begin{scope}[xshift=7.cm] 
\node[font=\small\bfseries] (titleB) at (0, 0.8) {(b) CosmoEvolve Virtual Lab};

\node[draw, thick, fill=purple!10, rounded corners, minimum width=5.2cm, minimum height=1.5cm] (pi) at (0, -0.3) {};
\node[font=\bfseries\tiny] at (0, 0.2) {Principal Investigator (PI)};

\node[draw, fill=white, font=\tiny, minimum width=1.4cm, minimum height=0.4cm] at (-1.6, -0.2) {Tools};
\node[draw, fill=white, font=\tiny, minimum width=1.4cm, minimum height=0.4cm] at (0, -0.2) {Skills};
\node[draw, fill=white, font=\tiny, minimum width=1.4cm, minimum height=0.4cm] at (1.6, -0.2) {Memory};
\node[font=\tiny, color=black!70] at (0, -0.7) {$\bullet$ Meetings $\bullet$ Assignments $\bullet$ Reviews};

\foreach \i/\x/\label in {1/-1.4/Student 1, n/1.4/Student $n$} {
    \node[draw, thick, fill=teal!5, rounded corners, minimum width=2.2cm, minimum height=1.2cm] (s\i) at (\x, -2.4) {};
    \node[font=\tiny\bfseries] at (\x, -2.0) {\label};
    
    \node[draw, fill=white, font=\tiny, minimum width=0.3cm] at (\x-0.7, -2.6) {T};
    \node[draw, fill=white, font=\tiny, minimum width=0.3cm] at (\x, -2.6) {S};
    \node[draw, fill=white, font=\tiny, minimum width=0.3cm] at (\x+0.7, -2.6) {M};
}
\node at (0, -2.4) {\dots};

\node[draw, thick, fill=blue!5, rounded corners, minimum width=5.5cm, minimum height=0.8cm] (shared) at (0, -4.1) {};
\node[font=\bfseries\tiny] at (0, -3.9) {Shared Lab State (Blackboard)};
\node[font=\tiny] at (0, -4.2) {$\bullet$ Discussion $\bullet$ Artifacts $\bullet$ Reviews $\bullet$ Skills};

\draw[->, thick, color=purple!80] (pi.south) -- ++(0,-0.2) -| (s1.north);
\draw[->, thick, color=purple!80] (pi.south) -- ++(0,-0.2) -| (sn.north);

\draw[->, thick, color=teal!80] (s1.south) -- ++(0,-0.2) -| (shared.170);
\draw[->, thick, color=teal!80] (sn.south) -- ++(0,-0.2) -| (shared.10);

\end{scope}

\end{tikzpicture}
\caption{System architectures of the two agentic frameworks. (a) \texttt{CMBEvolve} performs LLM-guided scientific code evolution through a typed tree search, with idea generation, branch selection, targeted mutation, execution, scoring, and score backpropagation. (b) \texttt{CosmoEvolve} simulates a virtual research lab in which a PI agent and student scientist agents, equipped with tools, skills, and memory, coordinate through a shared lab state.}
\label{fig:overall}
\end{figure}

\section{Introduction}

Human-driven discovery sometimes faces challenges in the modern era, including substantial time and monetary costs, potential cognitive biases, and increasing data complexity. Recent advances in artificial intelligence (AI), including large language models (LLMs) and AI agents, raise the possibility of moving beyond passive tools toward systems that can actively contribute to scientific workflows and even end-to-end discovery.


Recent studies have started to explore AI agents for cosmology and scientific discovery~\cite{laverick2024multiagentcosmologicalparameteranalysis,xu2025opensourceplanning,villaescusanavarro2025denarioprojectdeepknowledge}, with early demonstrations on realistic cosmological tasks such as weak-lensing analyses~\cite{borrett2026competingaiscientistsagentdriven}. Inspired by the work on self-evolving agents~\cite{novikov2025alphaevolve}, we discuss two complementary agentic AI systems that can be applied to scientific discovery in cosmology: one for quantitative tasks with explicit evaluation metrics, and one for more open-ended scientific workflows. We present our preliminary results on cosmological applications of our algorithms.


\section{Multi-agent systems for scientific discovery}
\label{sec:masintro}
We review the fundamental concepts of AI agents for autonomous scientific discovery. Formally, an agent $\mathcal{A}$ can be modelled as the tuple $\mathcal{A}=(\mathcal{L}, \mathcal{C}, \tau)$, where $\mathcal{L}$ is the LLM backbone, $\mathcal{C}$ is the context information received, and $\tau$ is the tools available for a single agent. Agents solve problems through multi-agent collaboration. Given a research problem and a dataset $\mathcal{D}$, the agentic system initiates a high-level goal $\mathcal{G}$, and then follows a policy $\pi$ to generate a sequence of actions $\left\{a_0, \ldots, a_N\right\}$ that execute a collection of sub-procedures aimed at advancing the system towards the solution. The policy may be viewed as a conditional probability distribution over actions given the high-level goal $\mathcal{G}$, the available tools $\mathcal{T}$, and the context $\mathcal{C}_{t-1}$ before the decision step $t$, such that
\begin{equation}
a_t \sim \pi(\cdot \mid \mathcal{G}, \mathcal{T}, \mathcal{C}_{t-1}).
\end{equation}

Autonomous scientific discovery may be formulated as the problem of finding a solution that maximizes the expected utility of a discovery trajectory. Let
\[
\Xi_{0:N}=\left(\mathcal{C}_0,a_0,\mathcal{O}_1,\mathcal{C}_1,\ldots,a_N,\mathcal{O}_{N+1},\mathcal{C}_{N+1}\right)
\]
denote the trajectory generated by the agentic system, where $\mathcal{O}_t$ represents the observation, tool output, or experimental result obtained during the discovery process. The system seeks a policy $\pi^\ast$, such that
\begin{equation}
\pi^\ast = \mathrm{arg\,max}_{\pi}\,
\mathbb{E}_{\Xi_{0:N}\sim\pi}\!\left[
U(\Xi_{0:N}; \mathcal{G}, \mathcal{D})
\right].
\end{equation}
Here, $U$ is a task-dependent utility function over the discovery trajectory. In quantitative settings, it may be defined by explicit metrics, such as parameter recovery or agreement with data. In more open-ended settings, it may instead reflect broader scientific value, such as generating hypotheses, identifying unexpected patterns, or proposing interpretable explanations.


\section{Methodology}
\label{sec:method}
We consider the two settings for agentic scientific discovery introduced in Sec. \ref{sec:masintro}: closed tasks with explicit quantitative objectives, and open-ended scientific workflows with only partially specified goals.

\subsection{\texttt{CMBEvolve}: automated algorithm discovery through LLM-guided tree search}
For tasks with explicit quantitative metrics, we use an algorithmic evolution approach to find the optimal solution. We present \texttt{CMBEvolve} \footnote{Will be made available publicly}, a package for automated scientific code evolution through LLM-guided tree search. Given a task for which the candidate solutions can be evaluated by a score $s$, we represent the search process as a rooted tree $\mathcal{T}=(V,E)$, where $V$ is the set of nodes and $E$ the parent-child edges. Each node belongs to one of the four types: task, idea generation and selection, code generation, and code mutation. Each node stores search statistics used by the selection rules, including best score $S^\ast$ and visit count $N$, together with the corresponding generated content and execution outputs. After code evaluation, the score assigned to a node is backpropagated from that node to the root, updating the statistics of all ancestor nodes. The general workflow of \texttt{CMBEvolve} is illustrated in Fig.~\ref{fig:overall}(a).

\subsection{\texttt{CosmoEvolve}: towards open-ended scientific discovery}
\texttt{CosmoEvolve} \footnote{Will be made available publicly} is a package designed for open-ended scientific discovery. It simulates a virtual research laboratory consisting of one principal-investigator (PI) agent and a community of student scientist agents. The PI agent acts at the supervisory level, observes a summary of the current lab state, then selects an action from a finite discrete action space, including group meeting, individual meeting, and task assignment. The student scientist agents then carry out the scientific work independently according to the PI's decisions. Each student can dispatch subtasks into its own specialized subagents, including data and file exploration, planning, and code implementation. In addition, \texttt{CosmoEvolve} is engineered around explicit skills, tools, and context management. Agents are constructed through role-specific instructions, a compact skill index, and persistent memory. Skills are loaded on-demand, and tool access is controlled using allowlists for each agent. The overall workflow of \texttt{CosmoEvolve} is illustrated in Fig.~\ref{fig:overall}(b).

\begin{figure}[ht]
\centering
\begin{minipage}{0.6\linewidth}
\centering
\includegraphics[width=\linewidth]{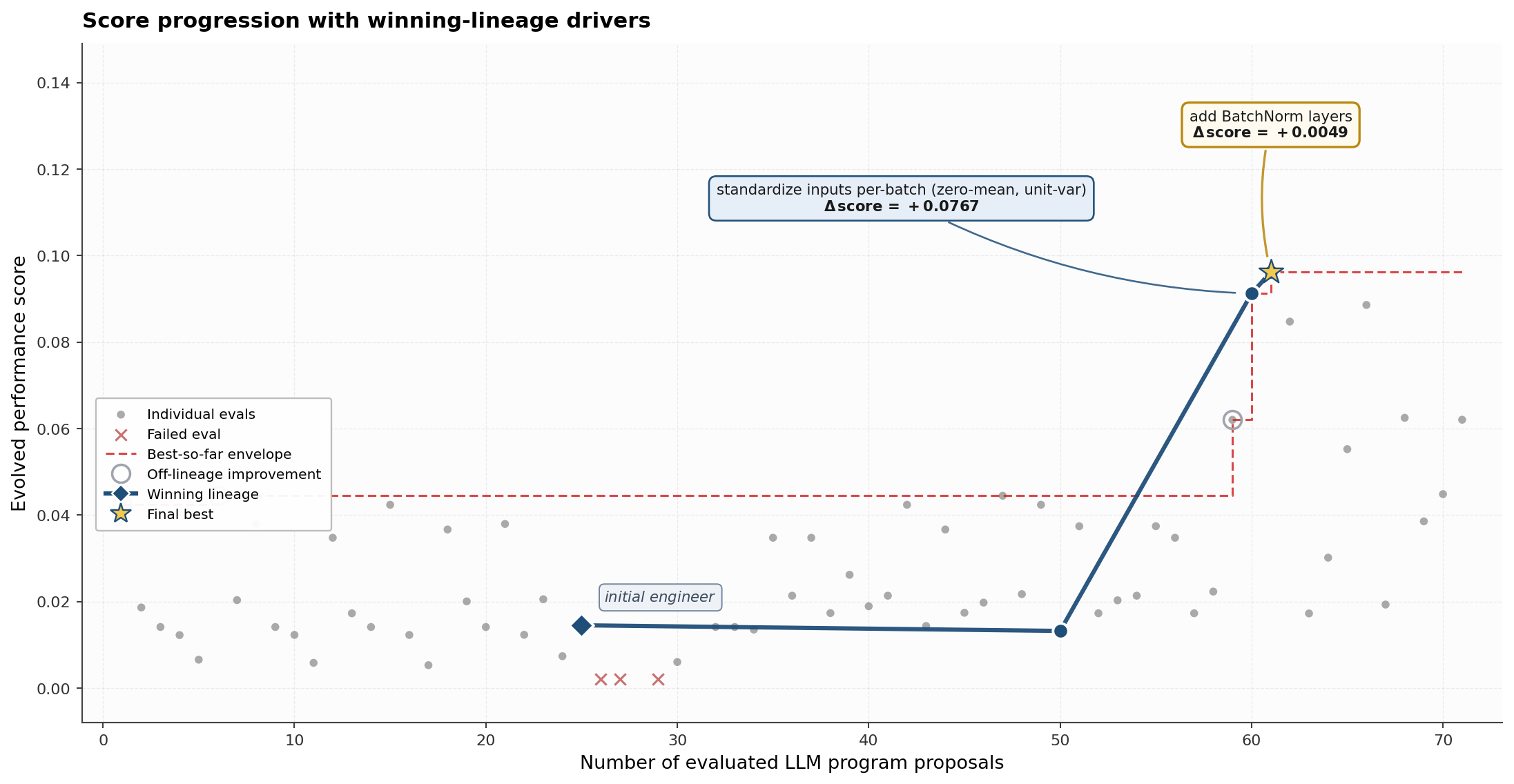}
\end{minipage}\hfill
\begin{minipage}{0.4\linewidth}
\centering
\includegraphics[width=\linewidth]{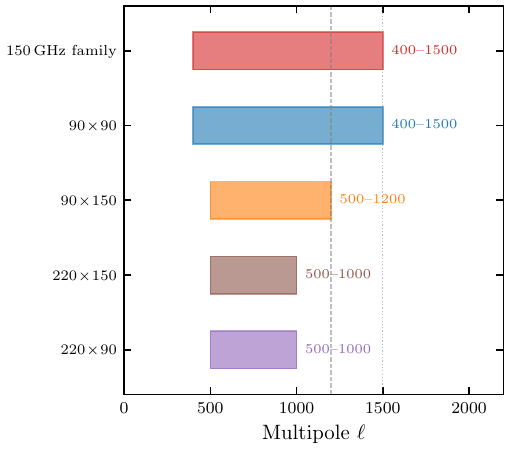}
\end{minipage}
\caption{\protect\textit{Left}: \protect\texttt{CMBEvolve} score evolution on the OoD task. \protect\textit{Right}: ACT DR6 pair-dependent diagnostic windows from \texttt{CosmoEvolve}, with reference cuts at \(\ell=1200\) and \(\ell=1500\). These are not ACT scale-cut recommendations; the official analysis supports multipoles up to $\ell\simeq5000$.}
\label{fig:results}
\end{figure}

\section{Preliminary Results}
\label{sec:results}
We present preliminary results on two cosmological tasks. First, we apply \texttt{CMBEvolve} to out-of-distribution (OoD) detection in weak-lensing maps, a quantitative task with a clear objective. Second, we apply \texttt{CosmoEvolve} to ACT DR6 data analysis, a more open-ended task involving iterative exploration and diagnostic refinement.

\subsection{Out-of-Distribution detection of weak lensing maps}
We consider the OoD detection task from the FAIR Universe Weak Lensing ML Uncertainty Challenge \footnote{\url{https://www.codabench.org/competitions/10902/}} as a benchmark for \texttt{CMBEvolve}. The inputs are simulated weak-lensing maps designed to mimic Hyper Suprime-Cam observations, with each map labelled by the cosmological parameters \(\Omega_m\) and \(S_8\), along with nuisance parameters describing baryonic and photometric-redshift systematics. We refer to Ref.~\cite{Dai:2026jme} for further details of the simulation setup, map-making procedure, and evaluation metric. The OoD benchmark is constructed from maps simulated with parameter distributions different from those used in the training set. The left panel of Fig.~\ref{fig:results} shows how the score improves over the number of iterations, illustrating how \texttt{CMBEvolve} improves performance iteratively through successive refinements of candidate solutions during tree search.


\subsection{ACT DR6 data analysis}

We also consider an open-ended ACT DR6 data-analysis task with \texttt{CosmoEvolve}, in which the system is given public ACT DR6 data products \footnote{\url{https://lambda.gsfc.nasa.gov/product/act/act_dr6.02/}} but no predefined scientific objective. It autonomously explored the released ACT DR6 temperature maps and produced beam-aware split-cross pseudo-\(C_\ell\) diagnostics, finding percent-level within-channel stability~\cite{PX:2604.00011}. A related multi-frequency coherence analysis~\cite{PX:2604.00012} identified pair- and scale-dependent behavior among channel combinations. These AI-generated diagnostic windows, shown in the right panel of Fig.~\ref{fig:results}, should be treated with much caution: the official ACT analysis goes far beyond this simplified study, with extensive null tests, mitigation strategies, and foreground modelling supporting the use of multipoles up to $\ell\simeq5000$. Related limitations were noted in Parallel OpenReview \footnote{\url{https://parallel-review-689836870161.us-central1.run.app/forum?id=2604.00012-R1}} and can inform future iterations of the agent harness, i.e. the system built around the LLM. These results show how \texttt{CosmoEvolve} can explore open-ended scientific problems, while highlighting the need for robust review and validation mechanisms for AI-generated diagnostics.



\section*{Acknowledgments}
We thank Boris Bolliet, Andy Nilipour, Francisco Villaescusa-Navarro, Pablo Villanueva-Domingo, and Íñigo Zubeldia for their support of this work. We also thank the \texttt{Cmbagent} team, the \texttt{Denario} team, and Erwan Allys’ group for helpful discussions and support. LX acknowledges support from the China Scholarship Council Cambridge Scholarship (grant number 202408060222). TB acknowledges support from PhD Studentship at the Infosys-Cambridge AI Centre. We also acknowledge Parallel Science \footnote{\url{https://parallelscience.org}}, including Parallel ArXiv \footnote{\url{https://papers.parallelscience.org}} and Parallel OpenReview \footnote{\url{https://reviews.parallelscience.org}}, f
or providing infrastructure for the dissemination and discussion of this work. We are also very grateful to ACTors for pointing to us an unfortunate framing of the AI generated ACT results in the first version of this preprint.

\section*{References}

\bibliography{moriond}

\end{document}